\documentclass{article}

\usepackage{arxiv}

\usepackage[utf8]{inputenc} 
\usepackage[T1]{fontenc}    
\usepackage{hyperref}       
\usepackage{url}            
\usepackage{booktabs}       
\usepackage{amsfonts}       
\usepackage{nicefrac}       
\usepackage{microtype}      
\usepackage{lipsum}
\usepackage{setspace}
\usepackage{layouts}
\usepackage{geometry}
\usepackage{graphicx}
\usepackage{amsmath}
\usepackage{float}

\newcommand\tab[1][0.5cm]{\hspace*{#1}}

\newcommand{\allocatepeer}[1][m \rightarrow n]{A_{#1,t}}
\newcommand{\netconsumption}[1][n]{p^{-}_{#1,t}}
\newcommand{\netproduction}[1][n]{p^{+}_{#1,t}}

\newcommand{\DirectedIncidence}{\mathcal{K}}
\newcommand{\InverseAPInjection}{\mathcal{J}}
\newcommand\diag[1]{\operatorname{diag}\left(#1\right)}

{}
{}
{}

\newcommand*{\email}[1]{\href{mailto:#1}{\nolinkurl{#1}} }

\begin{document}

\title{Carbon Leakage in a European Power System with Inhomogeneous Carbon Prices}

\author{
        Markus Schlott
        \thanks{\email{schlott@fias.uni-frankfurt.de}}\\
        Frankfurt Institute for Advanced Studies\\
	    Goethe Universität, FB Physik\\
	    Frankfurt am Main, Hessen, 60438, Germany\\
	\And
	    Omar El Sayed \\
        Frankfurt Institute for Advanced Studies\\
	    Goethe Universität, FB Physik\\
	    Frankfurt am Main, Hessen, 60438, Germany\\
	\And
	    Mariia Bilousova\\
	    Frankfurt Institute for Advanced Studies\\
	    Goethe Universität, FB Wirtschaftswissenschaften\\
	    Frankfurt am Main, Hessen, 60438, Germany\\
	\And
	    Fabian Hofmann\\
        Frankfurt Institute for Advanced Studies\\
	    Goethe Universität, FB Physik\\
	    Frankfurt am Main, Hessen, 60438, Germany\\
	\And
	    Alexander Kies\\
        Frankfurt Institute for Advanced Studies\\
	    Goethe Universität, FB Physik\\
	    Frankfurt am Main, Hessen, 60438, Germany\\
	\And
	    Horst Stöcker \\
        Frankfurt Institute for Advanced Studies\\
	    Goethe Universität, FB Physik\\
	    Frankfurt am Main, Hessen, 60438, Germany\\
	    GSI Helmholtzzentrum für Schwerionenforschung\\
	    Darmstadt, Hessen, 64291, Germany\\
}

\maketitle

\begin{abstract}
Global warming is one of the main threats to the future of humanity and extensive emissions of greenhouse gases are found to be the main cause of global temperature rise as well as climate change. During the last decades international attention has focused on this issue, as well as on searching for viable solutions to mitigate global warming. In this context, the pricing of greenhouse gas emissions turned out to be the most prominent mechanism: First, to lower the emissions, and second, to capture their external costs. By now, various carbon dioxide taxes have been adopted by several countries in Europe and around the world; moreover, the list of these countries is growing. However, there is no standardized approach and the price for carbon varies significantly from one country to another. Regionally diversified carbon prices in turn lead to carbon leakage, which will offset the climate protection goals. In this paper, a simplified European power system with flexible carbon prices regarding the Gross Domestic Product (GDP) is investigated. A distribution parameter that quantifies carbon leakage is defined and varied together with the base carbon price, where the combination of both parameters describes the spatially resolved price distribution, i.e. the effective carbon pricing among the European regions. It is shown that inhomogeneous carbon prices will indeed lead to significant carbon leakage across the continent, and that coal-fired electricity generation will remain a cheap and therefore major source of power in Eastern and South-Eastern Europe - representing a potential risk for the long term decarbonization targets within the European Union.
\end{abstract}

\newpage

\section{Introduction} \label{sec:sec1}

Global warming is a continuous increase of the observed average temperature on Earth, mainly caused by anthropogenic greenhouse gas emissions. To mitigate the negative effects of global warming and slow down the temperature rise, power systems around the world are being transformed towards high shares of renewable energy sources.\\
\tab Pricing the emissions of greenhouse gases, in particular carbon dioxide, has proven to be an efficient measure to include their external cost in economic processes. However, a wide variation in carbon prices around the globe will lead to carbon leakage, a geospatial displacement of energy generation from conventional sources. For instance, a decrease of carbon dioxide emissions due to a specific taxation in one country will raise the emissions in another one, where energy-intensive industries are moved around to minimize cost.

\subsection{The Evidence for Carbon Leakage}

Carbon leakage can occur via different channels \cite{kuik2009climate}. The pure energy channel is the most prominent one and describes any energy related infrastructure and its alteration under external pressure, e.g. shifting of international fossil fuel trade routes and the corresponding fuel consumption. Other channels are the political, economic and competitiveness channels. Carbon leakage has also been investigated in a number of research works, e.g. with regard to the EU Emissions Trading System (EU ETS):\\
\tab Verde \cite{verde2020impact} studied this relationship with a focus on the years 2005-2012. He did not find evidence for significant carbon leakage but raised potential caveats. However, it was not explored whether the EU ETS had long-term effects on the economy by causing investment leakage.\\
\tab Naegele and Zaklan \cite{naegele2019does} investigated changes in the trade flows, particularly in embodied carbon, and whether the EU ETS causes carbon leakage in the European manufacturing sector, but they did not find evidence for this. Nonetheless, the most prominent tools in use against carbon leakage are border adjustment and output-based allocations. Both reduce the financial burden faced by industries in order to abate emissions \cite{kiviranta2020magnitude}.\\
\tab Arroyo et al. \cite{arroyo2015carbon} studied the policies in Europe, China and the United States and concluded that carbon leakage is not a strong counter-argument against actions by pioneers that adopt ambitious climate targets early on.\\
\tab On the other hand, Hassler and Krusell \cite{hassler2012economics}, who used an integrated assessment model to investigate different policies regarding carbon leakage, assessed the welfare consequences of carbon taxes that differ across regions. They showed that taxes on oil producers can benefit climate, whereas taxes on oil consumers can not.\\
\tab In this context, Böhringer et al. \cite{bohringer2010global} discussed whether climate policies do impact not only the countries undertaking them but also any of their trade partners, and find that regional climate protection goals may lead to general welfare losses, while policies tailored to avoid the accompanying carbon leakage may not.

\subsection{Carbon Taxation in Europe}

Zhu et al. \cite{zhu2019impact} took a look at carbon taxation in a coupled electricity and heat system across Europe and concluded that a carbon dioxide tax is mandatory for decarbonization. On the other hand and based on ideas from transition theory, Patt and Lilliestam \cite{patt2018case} argued that carbon pricing is outdated and several other policies are required to further curb greenhouse gas emissions.\\
\tab In general, there are different ways to estimate appropriate carbon prices: Schwenk-Nebbe et al. \cite{schwenknebbe2021} optimized a European power system including carbon taxes as a result of regional diversified emission quota attributions and showed that an inhomogeneous carbon price layout is needed to ensure a high degree of decarbonization.\\
\tab Ackerman and Stanton \cite{ackerman2012climate} studied the "social cost of carbon" using the DICE model and found that ambitious scenarios reaching zero or net negative emissions by the end of the century require high prices such as 150 to 500 USD per ton of carbon dioxide.\\
\tab In Sweden for instance, the carbon dioxide tax has been a major instrument of climate policy since 1991, and it is considered highly effective \cite{carbontax,johansson2000carbon}. The level of the Swedish carbon tax rose five-fold, from around 25 EUR per ton of carbon dioxide in 1991 to around 110 EUR per ton in 2020 \cite{carbontaxes2020}. Still, the tax level for industry stood significantly below this standard level during the entire period.\\
\tab Besides Sweden, several other European countries have experimented with carbon taxes but struggled to introduce a uniform tax system across the different economic sectors \cite{andersen2010europe}. In this context, Weitzman \cite{weitzman2017voting} and Nordhaus \cite{nordhaus2019climate} pointed out that it is easier to negotiate a single carbon price than to set different limits on carbon emissions per country.\\
\tab As it stands today, the different carbon taxes among European countries - besides the Swedish one - remain at relatively low levels \cite{carbontaxes2020}. However, Bayer and Acklin \cite{bayer2020european} showed that even these low prices lead to a significant reduction of carbon dioxide emissions.

\subsection{Approaches within the Present Paper}

In this paper we focus on carbon leakage from an economic-technical point of view, i.e. as it rises from cost-minimizations within an inhomogeneous carbon pricing scheme. For this, we define effective carbon prices. They are assumed to be higher in wealthier regions than as compared to less wealthy ones, where the term "wealth" is reflected by the Gross Domestic Product (GDP) per capita.\\
\tab In order to study the impact of those regionally diversified carbon prices, a simplified European power system, where single countries are joined together into greater regions, is modelled and cost-optimized.\\
\tab Besides the overall network design, i.e. its grade of decarbonization and renewability as found by the optimization, carbon induced power flows in the transmission network are tracked and analysed. The tracking is performed with regard to general flow tracing methods as initially developed by Bialek et al. \cite{bialek1996tracing} in the 1990s. Such flow tracing methods have been adopted and widely used by the research community, e.g. for carbon accounting \cite{tranberg2019real}.\\
\tab It is finally shown that the applied pricing scheme causes carbon leakage, and that this leakage effect clearly endangers the ambitious emission reduction targets within the European Union.\\
\tab The model and the underlying data are open access and accessible to any interested reader. Both, model and data are archived on GitHub as the "EU11-CL" repository, a copy can be obtained via \url{https://github.com/carbon-leakage/EU11-CL}.\\
\tab This paper is a sophisticated version of our conference paper at the ICPES 2019 \cite{schlott2019carbon}, where rudimentary first insights on the discussed results were given.

\section{Methodology} \label{sec:sec2}

A simplified European power system with inhomogeneous carbon prices is investigated. The underlying model is built from scratch: It consists of loads attached to nodes and links, representing the European transmission network. It is subject to a greenfield investment problem and becomes cost-optimized. All statements in Ch. \ref{sec:sec3} and Ch. \ref{sec:sec4} will refer to the results from these cost-optimizations.

\begin{figure}[b]
    \centering
    \includegraphics[width=.47\textwidth]{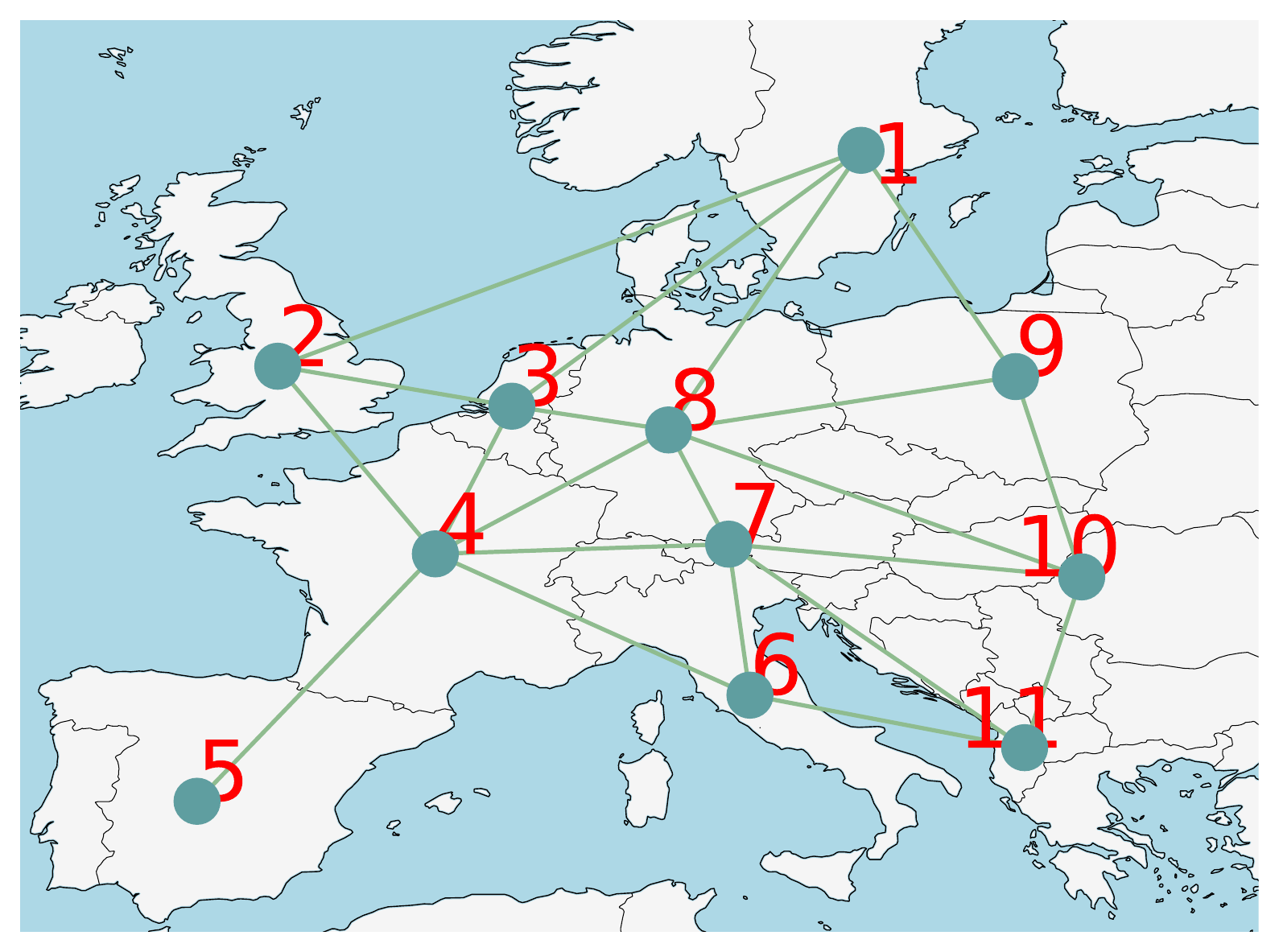}
    \caption{Topology of the investigated energy network. The clustering from individual European countries to eleven greater regions is performed with respect to the national GDP per capita, whereby neighbouring countries with a comparable GDP are joined together into single nodes as described in Tab. \ref{tab:nodes}.}
    \label{fig:topology}
\end{figure}

\begin{table*}[b]
\begin{center}
\resizebox{.89\textwidth}{!}{    \begin{tabular}{||c|c|c|c|c|c|c||}
        \hline Node  & Countries & Region & Short & GDP per capita [mu] & Population [m] & $\left<d_{n}\right>$ [GW] \\
        \hline
        \hline 1 & SE/NO/DK/FI & Scandinavia & SC & 60,149 & 26.8 & 45.4\\
        \hline 2 & IE/GB & Britain & GB & 44,869 & 71.3 & 42.1\\
        \hline3 & NL/BE/LU & Benelux & BE & 51,745 & 29.3 & 23.8\\
        \hline 4 & FR & France & FR & 41,463 & 67.0 & 54.3\\
        \hline5 & ES/PT & Iberia & IB & 29,193 & 57.0 & 34.8\\
        \hline6 & IT & Italy & IT & 34,318 & 60.4 & 36.8\\
        \hline 7 & AT/CH & Alps & AL & 66,877 & 17.4 & 15.0\\
        \hline 8 & DE & Germany & DE & 48,195 & 82.9 & 59.1\\
        \hline 9 & PL/LT/LV/EE & Baltic & BC & 15,998 & 44.0 & 21.6\\
        \hline 10 & RO/BG/HU/CZ/SK & Eastern & EA & 15,494 & 52.3 & 26.2\\
        \hline 11 & GR/SI/HR/RS/AL/BA/ME/XK/MK & Balkans & BK & 12,842 & 34.6 & 17.5\\
        \hline
    \end{tabular}}
    \caption{Nodes and their corresponding regions. The GDP per capita is given in monetary units [mu]. GDP and population data were obtained from the Open Data Platform of the World Bank \cite{worldbankdata}. The value $\left<d_{n}\right>$ denotes the average hourly electricity demand of a region as obtained from the Open Power System Data Project \cite{opsd}. All data refer to the year 2018.}
    \label{tab:nodes}
\end{center}
\end{table*}

\begin{table*}[b]
\begin{center}
\resizebox{.69\textwidth}{!}{    \begin{tabular}{||c|c||c|c||c|c||}
        \hline Link & Capacity [MW] & Link & Capacity [MW] & Link & Capacity [MW] \\
        \hline
        \hline IB-FR & 8000 & SC-DE & 6715 & AL-EA & 2200\\
        \hline FR-BE & 4300 & SC-BC & 2300 & DE-BC & 3000\\
        \hline FR-GB & 5400 & BE-DE & 8300 & DE-EA & 2600\\
        \hline FR-IT & 4350 & IT-AL & 7895 & BC-EA & 1590 \\
        \hline FR-AL & 3700 & IT-BK & 1880 & EA-BK & 6378\\
        \hline FR-DE & 4800 & AL-DE & 12200 & BE-GB & 2000\\
        \hline SC-GB & 1400 & AL-BK & 1200 & SC-BE & 1400\\
        \hline
    \end{tabular}}
    \caption{Link Capacities in [MW] according to the TYNDP report \cite{tyndp} from 2016.}
    \label{tab:links}
\end{center}
\end{table*}

\subsection{Topology}

The simplification from individual European countries to eleven greater regions is performed by clustering, where neighbouring countries with a similar GDP per capita are joined together. Then each cluster represents one node as shown in Fig. \ref{fig:topology}. More information on these nodes are given in Tab. \ref{tab:nodes}.\\
\tab Nodes are interconnected via high voltage, active transmission links. The capacity of each link was derived from ENTSO-E's Ten-Year Network Development Plan (TYNDP) report \cite{tyndp} from 2016 and is given in Tab. \ref{tab:links}, where in-country connections are aggregated according to the network topology as shown in Fig. \ref{fig:topology}.\\
\tab Then the 21 links from Tab. \ref{tab:links} have a combined transmission capacity of approximately 92 GW. This value lies slightly above the actual combined values of transmission between the countries in the high voltage transmission grid of Europe, which amounts to roughly 80 GW. The deviation represents scheduled expansions of the transmission grid in the medium-term future.\\
\tab Note that the power system model makes use of active link branches instead of passive line elements and therefore ignores the Kirchhoff's Voltage Law. This approach was taken in order to retrace and analyse the mean power flows within a simple transport model. The disadvantages of this method are discussed at the end of this chapter.\\

\subsection{Carbon Pricing}

\begin{figure}[b]
    \centering
    \includegraphics[width=.84\textwidth]{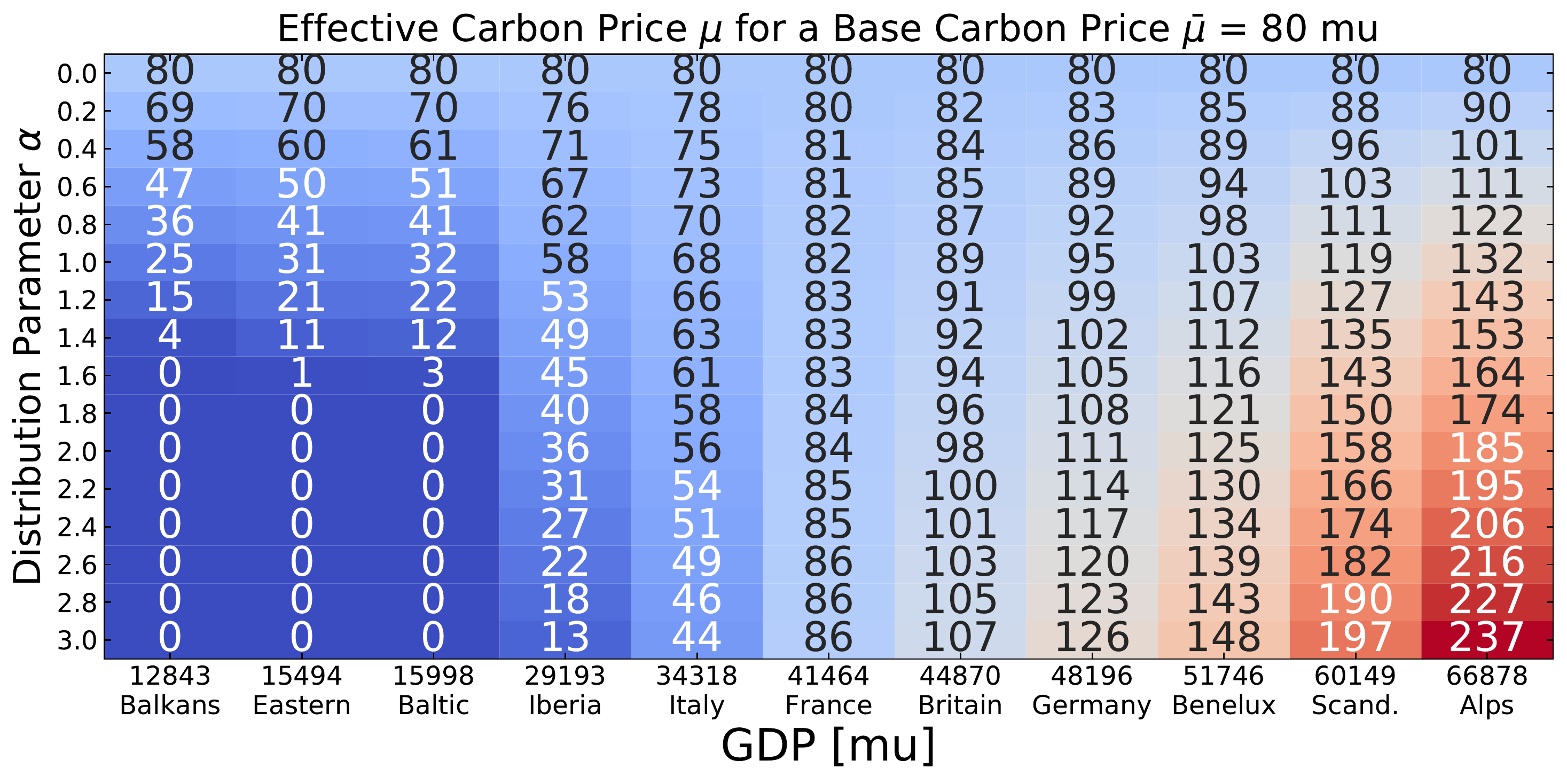}
    \caption{Example of effective carbon prices per region (color scale) for a base carbon price of $\bar{\mu} = 80$ mu and dependent on the distribution parameter $\alpha$. Blue and red tones denote low and high prices respectively. For inhomogeneous distributions ($\alpha \geq 1.6$), carbon prices fall drastically and eventually vanish in regions with small GDPs (Baltic, Eastern and Balkans), while prices strongly increase in regions with comparably high GDPs (Alps, Scandinavia and Benelux).}
    \label{fig:price_map}
\end{figure}

Carbon prices vary between regions. Therefore, regions with low local carbon prices possess financial advantages as compared to regions with higher ones, which means that they can rely on carbon intensive energy at comparably low cost. These price differences lead to the systemic carbon leakage effect as stated in the introduction: The shift of conventional energy generation from one geographical region to another. It gets amplified by the degree of price inhomogeneities among the regions, where the greater the price spread, the bigger the leakage effect.\\
\tab In this work, local effective carbon prices $\mu$ are defined as a function of the GDP per capita and the systemic base carbon price $\bar{\mu}$:

\begin{align} \label{eq:carbon_price}
\begin{split}
    \mu\left(\text{GDP}\right) &= \bar{\mu} + \alpha \bar{\mu} \frac{\text{GDP}}{\text{GDP}_{\emptyset}} - \alpha\bar{\mu}\\
    &= \bar{\mu}\left(1+\alpha\left(\frac{\text{GDP}}{\text{GDP}_{\emptyset}}-1\right)\right)\\
    \mu\left(\text{GDP}\right) &\geq 0 \text{,}
\end{split}
\end{align}

where $\text{GDP}_{\emptyset} = \sum_n \text{GDP}_n \left<d_n\right> / \sum_n \left<d_n\right>$ is the demand-weighted average GDP per capita. It defines the relative price scale by weighting the GDPs according to the average hourly electricity demands $\left<d_n\right>$ as given in Tab. \ref{tab:nodes}.\\
\tab The distribution parameter $\alpha$ quantifies the distribution of carbon prices among the regions and represents the slope of the function: The higher $\alpha$, the higher the carbon price inhomogeneities and the greater the price spread between the regions. The value of the distribution parameter $\alpha$ as well as the base carbon price $\bar{\mu}$ vary throughout this paper, while all absolute cost values are given in monetary units [mu].\\
\tab For an $\alpha <  \left(1 - \text{GDP}^\text{min}_n / \text{GDP}_{\emptyset} \right)^{-1}$ the second condition is trivially true. However, if $\alpha$ is chosen above this maximum threshold, the demand-weighted carbon price does not equal the carbon base price anymore. Furthermore, since negative prices are not considered, the average carbon price over all individual nodes is not constant but grows by increasing $\alpha$.\\
\tab With respect to a variation of $\alpha$ and a fixed base carbon price of $\mu = 80$ mu Fig. \ref{fig:price_map} exemplifies how the effective carbon price is distributed among the regions. It shows that - for strongly inhomogeneous distributions ($\alpha \geq 1.6$) - carbon prices do vanish in regions with small GDPs (Baltic, Eastern and Balkans), whereas they strongly increase in regions with comparably high GDPs (Alps, Scandinavia and Benelux).\\
\tab This is the main feature of Eq. \ref{eq:carbon_price}. Later on all results from the optimization procedure will be discussed with regard to this bidirectional behaviour.

\subsection{Model Description}

The European power system described above is subject to a linear scheme which minimizes total system cost. The optimization objective is to find the most cost-efficient solution, i.e. the cost-optimal system configuration under various physical, technical and economical constraints.\\
\tab After the optimization all power flows within the network are tracked from source to sink, and the effective carbon consumption of each region is determined by assigning carbon dioxide emissions to all carbon charged flows. This power flow tracing will clearly display the severe issue of carbon leakage in Europe.

\subsubsection{Optimization Scheme}

The complete optimization objective reads

\begin{align} \label{eq:minimization}
 \min_{g,G,F} & \left(\sum_{n,s} c_{n,s} \cdot G_{n,s} + \sum_{n,s,t} o_{s} \cdot g_{n,s,t} \right) \text{.}
\end{align}

It consists of capital cost $c_{n,s}$ for the installed capacity $G_{n,s}$ of any energy carrier $s$ at node $n$ and marginal generation cost $o_{s}$ for the energy generation $g_{n,s,t}$ of any carrier type $s$ at the same node $n$ and time $t$, as given in Tab. \ref{tab:cost_assumptions}.\\
\tab Note that carbon prices are reflected by an increase in marginal generation cost of any conventional energy carrier included.\\
\tab At every node, generation capacities of different energy carriers can be built up and dispatched. The model includes wind turbines (wind), solar photovoltaic modules (solar PV), coal power plants (coal) and open cycle gas turbines (OCGT or simply gas).\\
\tab Storage units are considered as well. They are given as conventional dammed-hydro reservoirs (hydro), as well as expandable battery units (battery) and provide additional energy storage capabilities.\\
\tab Note that pumped-hydro reservoirs are not considered, and that the expandable energy carriers were chosen to draw a simple duality between a renewable and a conventional infrastructure.\\
\tab The dispatch interaction between all of those generators and storage units, as well as between the different power flows on each link has to satisfy the given hourly demand at every node and time step over the full optimization span, the so-called nodal power balance:

\begin{align} \label{eq:power_balance}
\sum_s g_{n,s,t} - d_{n,t} = \sum_l K_{n,l} \cdot f_{l,t} \quad \forall \quad n \text{,} t \text{.}
\end{align}

Here, $d_{n,t}$ is the demand at node $n$ and time $t$;  $K_{n,l}$ is the network's incidence matrix, describing the relative network topology, and $f_{l,t}$ is the actual power flow on a link $l$ at time $t$.

However, the power dispatch from any generator is constrained:

\begin{align} \label{eq:disptach_constraint}
{g}^-_{n,s,t} \cdot G_{n,s} \leq g_{n,s,t} \leq {g}^+_{n,s,t} \cdot G_{n,s} \quad \forall \quad n \text{,} t \text{.}
\end{align}

It is limited by its minimum generation potential ${g}^-_{n,s,t}$ and by its maximum generation potential ${g}^+_{n,s,t}$, where $G_{n,s}$ describes the capacity of a carrier type $s$ at node $n$.\\
\tab Storage units obey additional constraints as given by the State-Of-Charge (SOC) equations, constraining their charging and discharging behaviour:

\begin{align} \label{eq:state_of_charge}
\begin{split}
\mathrm{SOC}_{n,s,t} &= \eta_0 \cdot \mathrm{SOC}_{n,s,t-1} + \eta_1 \cdot g_{n,s,t,\textrm{store}} - \eta_2^{-1} \cdot g_{n,s,t,\textrm{dispatch}}\\ &+ \mathrm{inflow}_{n,s,t} - \mathrm{spillage}_{n,s,t} \quad \forall \quad t > 1 \text{.}
\end{split}
\end{align}

Charging efficiencies are described by $\eta$, where $\eta_0$ represents standing losses (which are assumed to be small and hence neglected in the following), $\eta_1$ stands for efficiencies of electrical storage uptake and $\eta_2$ for efficiencies of electrical storage dispatch (both are assumed to be at 90\% for battery storage). The term $\mathrm{inflow}$ covers any external inflow, e.g. water inflow into hydro reservoirs and $\mathrm{spillage}$ stands for overflow losses, e.g. by reaching the maximum physical storage capacity.\\
\tab At last, all active power flows $f_l$ on a transmission link $l$ are constrained by the maximum transmission capacity $F_l$ of the considered link:

\begin{align} \label{eq:flows}
|f_l\left(t\right)| &\leq F_l \quad \forall \quad l \text{.}
\end{align}

To perform the optimization, we use the software toolbox Python for Power System Analysis (PyPSA) \cite{brown2017pypsa}. PyPSA has been widely used throughout the research community, especially in the field of renewable energies, e.g. to study the interplay of sector coupling and transmission grid extensions \cite{brown2018synergies}, or the impact of climate change on a future European power system \cite{schlott2018impact}.

\subsubsection{Power Flow Tracing}

Physical power flows throughout the network are evaluated in terms of flow tracing, as initially described by Bialek et al. \cite{bialek1996tracing}. Here, we refer to an algorithm developed within the framework of PyPSA written by F. Hofmann et al. \cite{hofmann2019}. We restrict our analysis to a specific flow tracing method, called Average Participation.\\
\tab In this scheme, all excess power at a node is assumed to get redistributed among the attached network connections in the proportion all incoming partial flows take compared to the total excess power of the considered node. The redistributed power flow is in turn allocated to the initial producers. This happens for each carrier type individually, which allows tracking of electricity from a power plant to a power consumers, i.e. from any energy source to any energy sink within the network.\\
\tab In matrix representation, the power flow allocation regarding the Average Participation scheme is based on the inverse injection matrix $\InverseAPInjection_{m,n,t}$. It is given by

\begin{align}
\InverseAPInjection_{m,n,t} = \left( \diag{p_{n,t}^+} + \DirectedIncidence_{m,l}^- \diag{f_{l}}_t \ K_{l,n} \right)^{-1} \text{,}
\end{align}

where $f_{l}$ is the power flow matrix, $K_{n,l}$ is the incidence matrix and $\DirectedIncidence_{m,l}^-$ is the negative part of the flow directed incidence matrix $\DirectedIncidence_{m,l} \left(t\right) = \text{sign} \left( f_l \right)_t K_{m, l}$. Then the peer-to-peer allocation $\allocatepeer$ for any time step $t$ is given by

\begin{align}
\allocatepeer = \InverseAPInjection_{m,n,t} \ \netproduction[m] \ \netconsumption \text{.} 
\end{align}

Here $\netproduction[m]$ labels all network injections at a node $m$, while $\netconsumption$ classifies any network related consumption at a node $n$, both at time $t$.\\
\tab All flows originating from conventional carrier types are tracked and their corresponding carbon entry to the atmosphere is determined. This is done at consumer level, i.e. at nodal sinks $n$. Then at each of these nodal sinks the corresponding carbon dioxide allocation in metric tons [t] is given as

\begin{align}
A_{n,s}^{\text{CO$_2$}} = \sum_{m,t,s} \allocatepeer \cdot e_s \text{,}
\end{align}

where $e_s$ denotes the power related carbon dioxide emissions from any carrier type $s$, i.e. coal and gas respectively, as given in Tab. \ref{tab:cost_assumptions}. A similar study of consumption based carbon emissions within the actual European power grid was recently carried out by Schäfer et al. \cite{schaefer2020}.

\subsection{Data and Assumptions}

The presented power system model is based on several datasets and assumptions:

\begin{itemize}
    \item \textbf{Temporal Resolution:} The total time span of the optimization was chosen to reflect one calendrical year in an hourly resolution.
    \item \textbf{Load Data:} The electricity consumption data for the eleven regions were obtained from the Open Power System Data Project \cite{opsd} for the reference year 2018. They represent any electrical load appearing on the left hand side of Eq. \ref{eq:power_balance}.
    \item \textbf{Generation Potentials:} The generation potentials of the different energy carrier types included, i.e. wind, solar PV, hydro, coal and gas, are determined by several approaches:
    \begin{itemize}
        \item \textbf{Conventional Carriers:} Generation from conventional energy carriers, i.e. coal and gas, is assumed to be fully dispatchable and does not rely on the input weather data in any way.
        \item \textbf{Renewable Carriers:} In order to determine the energy output from renewable energy carriers, weather data are taken into account. Regional weather datasets are processed to generation potentials by invoking the corresponding energy carrier type, e.g. a wind turbine or a solar panel model.
    \end{itemize}
    \item \textbf{Weather Datasets:} The relevant renewable generation potentials for wind and solar PV were obtained from the renewables.ninja database \cite{pfenninger2016renewables}, both times for the reference year 2013. The renewables.ninja dataset is one of the most popular datasets in the field of energy system applications \cite{kiescomparison}. Note that reanalysis weather data are commonly used to study renewable power systems \cite{jurasz2020review}.
    \begin{itemize}
        \item \textbf{Wind:} Wind potentials are based on wind speeds from the MERRA-2 reanalysis \cite{gelaro2017modern} with respect to the actual wind turbine fleets in the eleven regions.
        \item \textbf{Solar PV:} Solar PV potentials are based on the CM-SAF's SARAH satellite dataset \cite{muller2015surface} with default solar panel configurations.
        \item \textbf{Hydro:} Water inflow into hydro reservoirs as well as the reservoir's capacities were obtained from the RESTORE 2050 project \cite{alexander_kies_2017_804244} for the reference year 2011. All feasible hydro locations are assumed to be fully exploited, not extendable and hence set to a fixed value. The amount of energy stored in water inflow was modelled by invoking the Newtonian gravitational potential with respect to geographical height and surface runoff data \cite{kies2016restore}.
    \end{itemize}
    \item \textbf{Cost Assumptions:} Cost assumptions for the considered technologies were taken from different sources \cite{carlsson2014etri,schroeder2013current} and are annualised with a discount rate of 7\%. All assumptions are given in Tab. \ref{tab:cost_assumptions}.
    \item \textbf{Remark:} Note that total system cost are relatively robust with regard to the chosen input weather data but also to moderate variations in cost as shown by Schlachtberger et al. \cite{schlachtberger2018}.
\end{itemize}

\begin{table*}[b]
\begin{center}
\resizebox{.49\textwidth}{!}{\begin{tabular}{ ||l||r|r|r|| }

        \hline Technology & Capital Cost & Marginal Cost & Emissions \\
                      &  [mu/GW/a] & [mu/MWh] & CO$_2$ [ton/MWh] \\ 
        \hline
        \hline wind & 127,450 & 0.01 & 0\\
        \hline solar PV & 61,550 & 0.01 & 0 \\
        \hline hydro & 0 & 0 & 0\\
        \hline coal & 145,000 & 25 & 1 \\
        \hline OCGT/gas & 49,400 & 58.385 & 0.635\\
        \hline battery & 120,389  & 0 & 0\\
        \hline
    
    \end{tabular}}
    \caption{Annualised cost assumptions as derived from different sources \cite{carlsson2014etri,schroeder2013current}. All energy quantities refer to exergy values.}
    \label{tab:cost_assumptions}
\end{center}
\end{table*}

\subsection{Limitations}

The model has a limited scope as discussed in the follwoing:

\begin{itemize}
\item Countries were aggregated to regions by referring to neighbourhood and the GDP per capita, which limits the geographical resolution. Hörsch et al. \cite{hoersch2017} argued that these issues should not affect the quality of the results too drastically. Of course, a finer network topology could strengthen the significance of the results and enhance the micro-spatial precision.
\item The energy and storage carrier types wind, solar PV, coal, gas, hydro and battery were chosen to draw a simple duality between renewable and conventional infrastructures. An enhanced model would include a greater variety of additional carrier and storage types, e.g. nuclear power and pumped-storage hydroelectricity. Also no near future technologies such as carbon capture and storage, demand side management, or power-to-x conversions with respect to hydrogen and synthetic methane are considered.
\item Second order effects regarding costs were neglected. In reality, capital cost of the different carrier types, especially the ones from renewable sources, do vary substantially from one region to another. This would interfere with the number and capacity of deployed plants in the different regions as determined by the optimization. It may also favor conventional generation capacities over renewable ones.
\item No realistic time-dependent transformation scenarios, e.g. coal phase out plans or other political efforts to achieve a general shift towards fully renewability, were taken into account. The model rather focuses on coal power plants which are built up due to geospatial displacements caused by carbon leakage, especially in those regions that already provide large amounts of energy from coal \cite{coalexittracker}. The results demonstrate that inhomogeneous carbon prices will prolong this status quo, and hence strengthen the conventional infrastructure of these countries in the medium term future.
\item The calculated power flow is based on a transport model, i.e. the nodal power balance from Eq. \ref{eq:power_balance}, where the nodal power input and output have to be balanced at each time step. This approximation does not respect the Kirchhoff Voltage Law, but it allows to circumvent a non-linearity of the optimization problem. Since the model only comprises the aggregated high-voltage levels of the European transmission grid (high inertia and small voltage angle differences) the effect of the approximation can be seen as insignificant.
\end{itemize}

\subsection{Outlook}

Work is in progress on the following issues:

\begin{itemize}
\item As mentioned in the limitations, capital cost for the different carrier types are treated as equal in different countries. Schyska et al. \cite{schyska2020regional} have recently shown that a variation of capital cost for investments in renewables across the European countries (which vary strongly because of large economic and political differences) have a profound impact on the distribution of renewables. It also is a reasonable assumption that inhomogeneous carbon prices and inhomogeneous capital cost do partially oppose each other. The complex interaction of these cost effects could be studied in a future work.
\item The electricity sector as discussed here is only one part of the entire energy system. It consumes about a quarter of the primary energy demand of Europe. The analysis of carbon leakage in a sector-coupled energy model, where sectors might or might not be fully electrified, will be studied in future works. Brown et al. \cite{brown2018synergies} showed that such systems are easier to decarbonize if the flexibility potentials of the heating and transport sectors are fully utilized.
\item In contrast to the all-electric world considered in this paper, power-to-x technologies are often presented as alternatives. The major energy carriers in power-to-x scenarios are hydrogen and synthetic methane, produced via electrolysis and Sabatier or Fischer-Tropsch processes respectively. These technologies offer additional flexibility, especially if the heating and transport sectors are included.
\item In the model presented, the energy system is built up from scratch (greenfield investment problem). However, taking the existing power plant park into account (brownfield investment problem) and performing a pathway optimization will lead to different results: In such a setting, decommissioning dates of power plants and limited investment budgets per time period, e.g. per annum, must be considered as well. This allows for investigations of policy control mechanisms and a modelling of the interactions between policies, the energy system as well as decarbonization targets.
\end{itemize}

\section{Results} \label{sec:sec3}

The European power system as described in Ch. \ref{sec:sec2} is cost-optimized according to the minimization procedure as stated in Eq. \ref{eq:minimization}, with boundary conditions as defined by Eqs. \ref{eq:power_balance} - \ref{eq:flows}. Note that effective carbon prices $\mu$ are applied to total cost in terms of regional marginal cost for energy generation from conventional sources via Eq. \ref{eq:carbon_price}.\\
\tab In total 1616 $\bar{\mu}$-$\alpha$-combinations are optimized: While the base carbon price $\bar{\mu}$ is varied from $\bar{\mu} = 0$ mu to $\bar{\mu} = 400$ mu in steps of $\Delta \bar{\mu} = 4$ mu, the distribution parameter $\alpha$ varies from $\alpha = 0.0$ to $\alpha = 3.0$ in steps of $\Delta \alpha = 0.2$.

\subsection{Regional Analysis}

\begin{figure}[b]
    \centering
    \includegraphics[width=.36\textwidth]{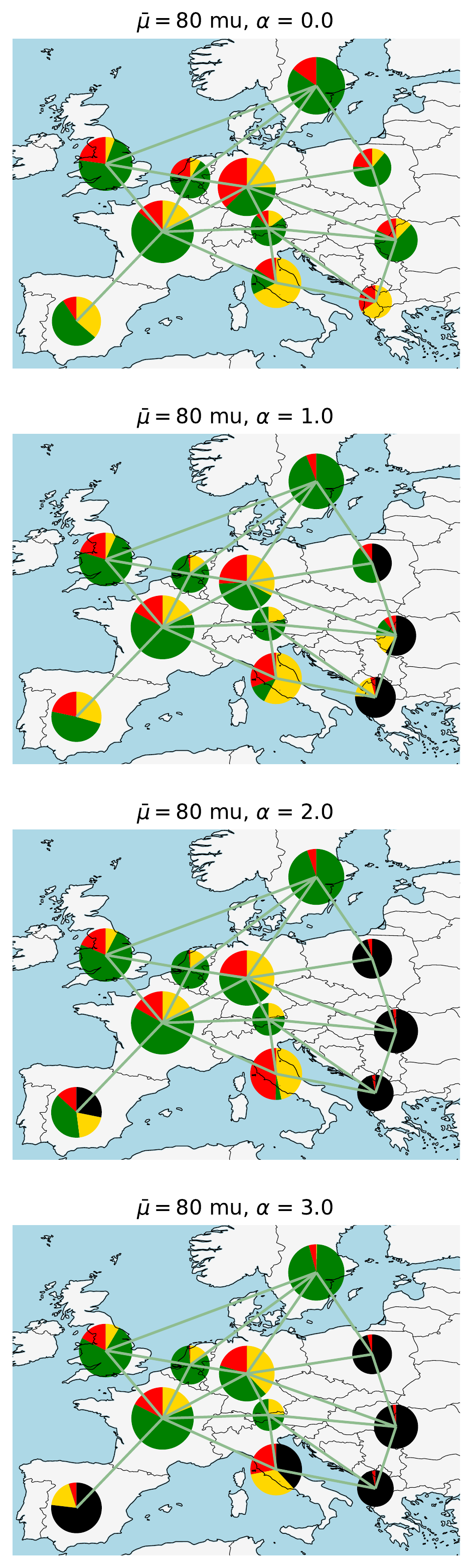}
    \includegraphics[width=.36\textwidth]{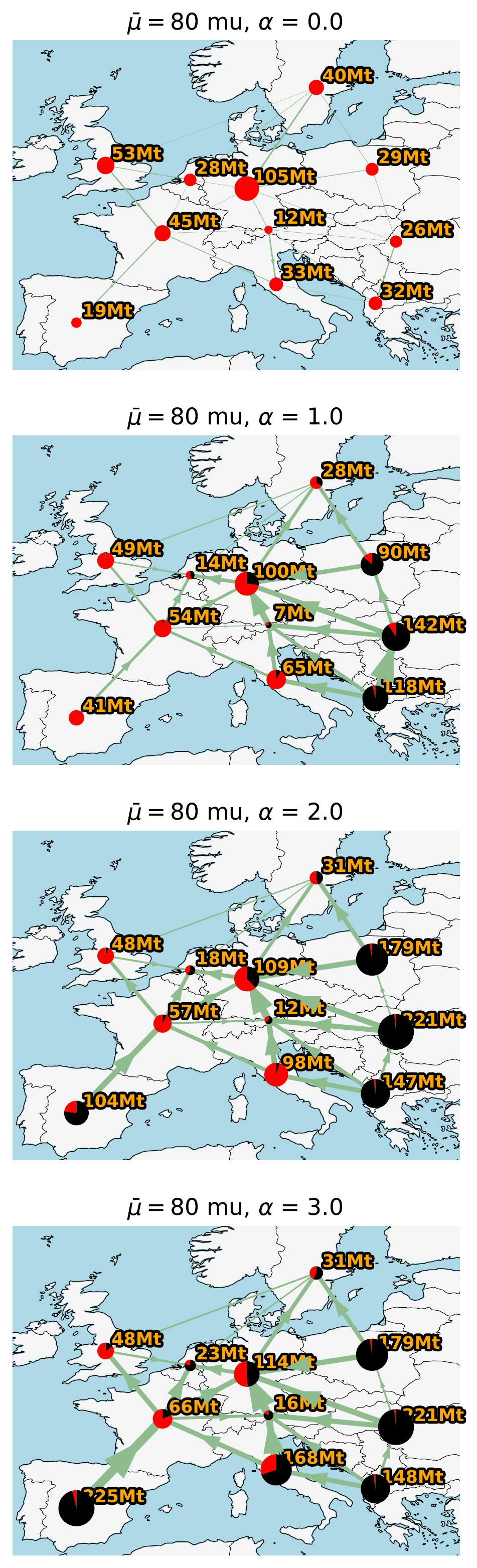}
    \caption{\textbf{Left:} Energy generation in [MWh] integrated over one year from wind (green), solar PV (yellow), coal (black) and gas (red). Circles indicate the overall energy generation within a region. Note, the higher $\alpha$, the more conventional capacities are deployed in regions with low effective carbon prices. \textbf{Right:} Mean carbon flow and mean carbon consumption integrated over one year, where the CO$_2$-flow per link (arrows) and the CO$_2$-consumption per node (circles) is given in metric tons [t], each time for the two energy carriers coal (black) and gas (red). Note, carbon leakage increases together with $\alpha$. Then cheap conventional energy does not only supply local loads but is also distributed within the network.}
    \label{fig:regional_analysis}
\end{figure}

First, we check where the optimization procedure deploys which generation capacities, how much energy is produced, how power flows on links are organised, and how carbon dioxide emissions can be attributed to consumers. The result of such assessment is illustrated in Fig. \ref{fig:regional_analysis}. Both sides present configurations for one single base carbon price $\bar{\mu} = 80$ mu, and four distinct values of the distribution parameter $\alpha$, i.e. $\alpha = \left[0.0, 1.0, 2.0, 3.0\right]$. The values are chosen to reflect the strongest appearance of carbon leakage within the $\bar{\mu}$-$\alpha$-solution space, while the general question on how the system evolves dependent on the base carbon price will be discussed in the subsequent section.\\
\tab Fig. \ref{fig:regional_cost_aspects} then shows how the regional cost aspects, i.e. the system costs, the consumer costs, the net profits, as well as the congestion rents of each link, evolve.\\
\tab Note that the selected base carbon price lies in between the most and second most expensive tax rate within the European Union, i.e. the Swedish and French taxes, which amount to around 110 EUR and 45 EUR per ton of carbon dioxide respectively \cite{carbontaxes2020}. If 1 EUR is identified with 1 mu, both the values for France and Scandinavia would be represented by a base carbon price $\bar{\mu} = 40$ mu and a distribution parameter $\alpha = 3.0$.

\subsubsection{Electricity Generation}

The left-hand side of Fig. \ref{fig:regional_analysis} shows the nodal energy generation integrated over a time span of one year. For $\alpha = 0.0$ the cost-optimization yields in a highly renewable energy system that consists mainly of wind (green) and solar PV (yellow) capacities. Also small and evenly distributed amounts of gas (red) capacities are required, which act as a relatively expensive backup to smoothen the renewable generation fluctuations rising from weather uncertainties.\\
\tab The renewable design is resulting from an evenly distributed and rather high base carbon price of $\bar{\mu} = 80$ mu. However, as $\alpha$ increases, an uneven distribution of effective carbon prices develops (see Fig. \ref{fig:price_map}). Then the system's preferred energy sources are shifting from renewables towards conventionals.\\
\tab For instance, when $\alpha$ reaches $\alpha = 1.0$, cost competing coal capacities (black) are getting deployed in regions with low effective carbon prices. This is the case for regions with comparably low GDPs, i.e. the Baltic, Eastern and Balkans regions.\\
\tab When the distribution parameter goes up to $\alpha = 2.0$, the effective carbon price is lowered further and eventually vanishes completely in the stated regions. Energy generation from coal becomes cheap and profitable again, and most of the local loads are supplied by conventional energy sources. Even the Iberian peninsula, a region with a relatively high GDP, starts to profit from this effect.\\
\tab At $\alpha = 3.0$, coal generation dominates not only the Iberian peninsula, but also the highly industrialised region of Italy starts covering a huge share of its energy needs by coal.\\
\tab Note that coal powered plants do not rely on weather conditions, and hence show no need for backup capacities. In regions with low and vanishing carbon prices backup capacities for the remaining renewable sources are more likely to be covered by coal instead of gas, too.\\
\tab Indeed, this dramatic shift to conventional carriers in regions with comparably low GPDs is the main source of the carbon leakage effect as explained in the introduction. It is analysed more closely in the following.

\subsubsection{Power Flow Tracing}

The right-hand side of Fig. \ref{fig:regional_analysis} shows the allocated carbon flows through the transmission network as well as their sources and their sinks at nodal level, both according to the power flow tracing method introduced in Ch. \ref{sec:sec2}.\\
\tab In doing so, the allocation first records the total power generation from coal and gas, and then tracks it from source to sink with regard to the chosen flow tracing method, i.e. Average Participation. In the next step, the allocated power is converted to CO$_2$-equivalents in metric tons [t] at sink as well as at link level, each time with respect to the energy related emissions as given in Tab. \ref{tab:cost_assumptions}. All flows are integrated over a time span of one year as covered by the optimization.\\
\tab Note that all carbon flows and carbon consumptions as given on the right-hand of Fig. \ref{fig:regional_analysis} are strongly correlated to the actual energy generation from conventional sources as given on the left-hand side respectively.\\
\tab Eventually, a complex systemic connection emerges: For a distribution parameter $\alpha = 0.0$ carbon charged flows are hardly present. Only moderate carbon dioxide emissions are ejected by gas acting as backup capacity.\\
\tab If the distribution parameter $\alpha$ increases, so the amount of energy generated from coal does. As seen on the left hand side of Fig. \ref{fig:regional_analysis}, for $\alpha = 1.0$ coal becomes so cheap that it dominates as source of electricity in the Baltic, the Eastern, as well as the Balkans regions. The cheap energy however does not only supply local loads but gets distributed among the network connections, mainly to the West and North, i.e. Germany and Scandinavia.\\
\tab This behaviour intensifies when the distribution parameter is further increased: For $\alpha = 2.0$ all network connections from east to west transmit an ever increasing amount of power from coal. The Iberian peninsula here does feed in a substantial amount of coal power into the transmission network and supplies France.\\
\tab For an $\alpha = 3.0$ carbon leakage reaches its maximum. Coal becomes a preferable source of power for all regions with a GDP below average, i.e. all regions left the middle regarding the abscissa as visible in Fig. \ref{fig:price_map}. Besides the Iberian peninsula, coal power is fed in by Italy and gets transmitted as far as link capacities allow for.\\
\tab As a leap ahead: The total aggregated energy generation from conventional sources with regard to the total dispatch is shown in Fig. \ref{fig:decarbonization}. The emissions displayed within the pie charts from Fig. \ref{fig:regional_analysis} are represented there by one respective line. $\alpha = 0.0$ is the bottom line, while $\alpha = 3.0$ is the top one. For a base carbon price of $\bar{\mu} = 80$ mu and for the lowest value of $\alpha = 0.0$, the total systemic energy supply from conventional sources amounts to roughly 10\%, and increases to about 40\% in the most extreme case, i.e. for $\alpha = 3.0$.

\subsubsection{Cost Aspects}

\begin{figure}[b]
    \centering
    \includegraphics[width=.88\textwidth]{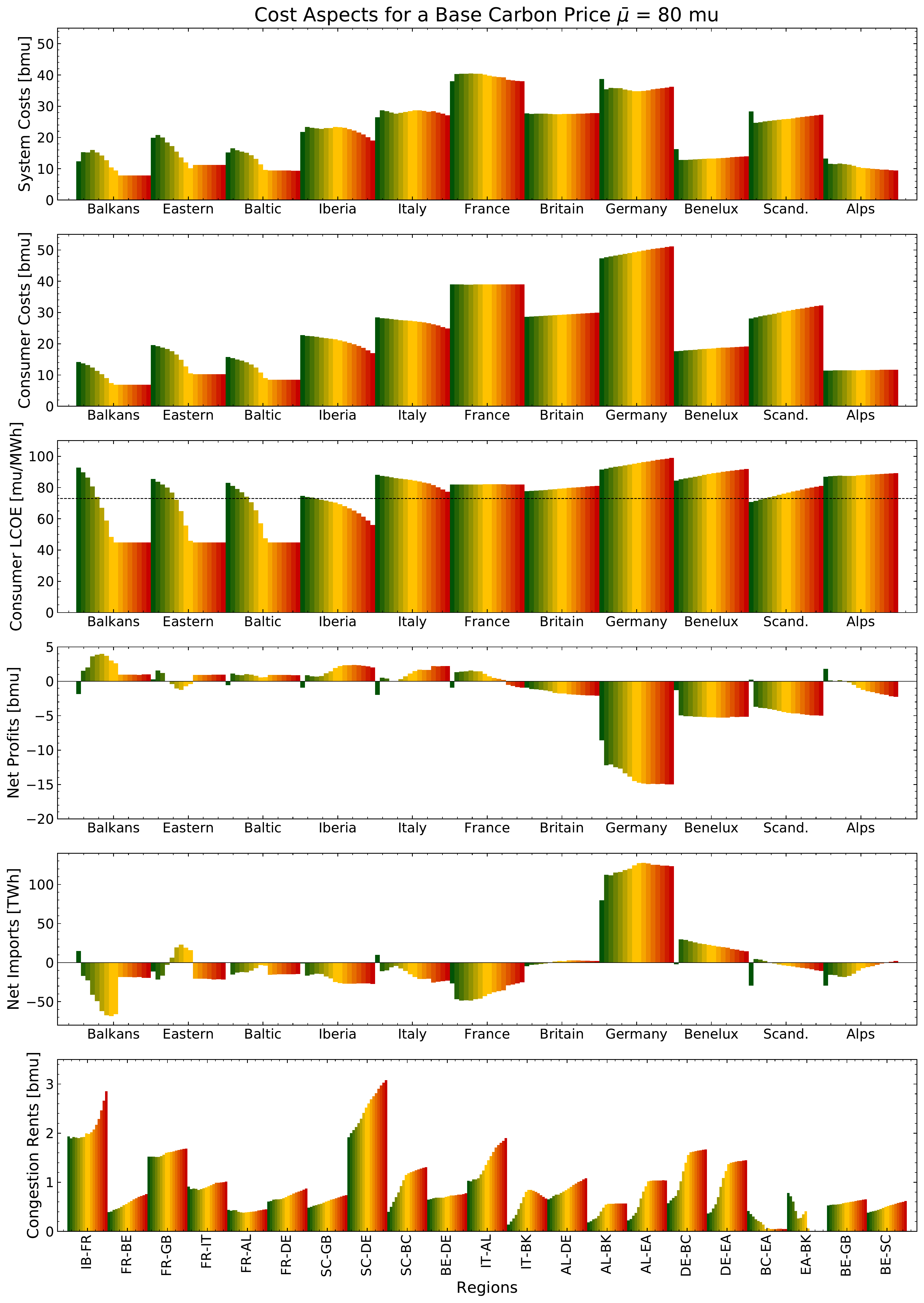}
    \caption{Regional cost aspects integrated over one year. The traffic light color scheme represents a rising distribution parameter from $\alpha = 0.0$ (dark green) to $\alpha = 3.0$ (dark red) in steps of $\Delta \alpha = 0.2$. Referred to the abscissa, carbon leakage lowers the consumer costs in all regions left the middle and vice versa increase them right the middle. Here, Germany acts as the main power importer and causes large congestions within the transmission network, especially for $\alpha \geq 1.6$, when coal power is imported from the East and South-East, but also Italy and the transit-region of Scandinavia.}
    \label{fig:regional_cost_aspects}
\end{figure}

Fig. \ref{fig:regional_cost_aspects} shows the different cost aspects with respect to each single region and link integrated over one year, each time for a base carbon price $\bar{\mu} = 80$ mu and a gradually rising distribution parameter $\alpha \in \left[0.0, 3.0\right]$ with steps of $\Delta \alpha = 0.2$, where the $\alpha$-interval is represented by a traffic light color scheme. Then an $\alpha = 0.0$, describing a highly renewable state, is represented by dark green, and an $\alpha = 3.0$, describing a more conventional state, by dark red. Note that each quantity of Fig. \ref{fig:regional_cost_aspects} must be compared $\alpha$-wise.\\
\tab The different cost aspects are: The regional system costs, i.e. the sum of all capital and marginal costs including carbon pricing, the regional consumer costs as given by the optimized shadow prices, the levelized cost of electricity (LCOE) at the corresponding consumer level, the net profits as given by the difference of system and consumer costs, the net imports and exports for comparison, as well as the congestion rents evoked by straining link capacities.\\
\tab We first recognize that the system costs in each region, i.e. the sum of all capital cost for carrier capacities as well as marginal cost of energy generation plus carbon pricing, are directly coupled to the distribution parameter. The higher $\alpha$, the larger this effect: Referred to the abscissa of Fig. \ref{fig:regional_cost_aspects}, all regions left the middle profit from an increasing amount of carbon leakage, whereas the regions right the middle are burdened by more and more expensive carbon taxes.\\
\tab The same is true on consumer level. The higher $\alpha$, the cheaper the conventional energy in regions on the left, but the higher the taxation in regions on the right.\\
\tab The cost at consumer level are also presented in terms of the consumer LCOE, obtained by dividing the consumer cost by the regional loads integrated over one year. It is remarkable that the consumer LCOE drops by half for the Baltic, the Eastern and the Balkans regions, while it increases by more than a quarter in Germany, Benelux and Scandinavia. For comparison, the dashed black line indicates the average LCOE with respect to the total system cost as taken over the whole of Europe. It is also displayed by one single grid entry in Fig. \ref{fig:lcoe}.\\
\tab The difference between the system cost and the consumer cost yields in the net profits. The net profits are in turn correlated to two other quantities: The net imports and exports as well as the congestions on transmission links.\\
\tab Here, several things are observable: First, the net profits show that most regions left the middle do not only profit from carbon leakage locally but also in terms of exports to regions right the middle. Especially Germany shows large net imports from its geographical surrounding.\\
\tab Second, the sum of all net profits does not equal zero. The remaining cost are given by congestion rents, which sharply rise together with $\alpha$. In the present case this means that traffic rises, and that the transmission system is strained more often, resulting in additional cost. It is also visible that mainly those links that connect to regions affected by carbon leakage are congested, e.g. the ones from Germany to the East, South and North, but also the link between Iberia and France. In these cases the congestion rents rise by a factor of two or even three.

\subsection{Systemic Analysis}

For a better understanding of how carbon leakage evolves with respect to the base carbon price $\bar{\mu}$ in combination with an increasing distribution parameter $\alpha$, we now present the overall system configuration across Europe, i.e. the total amount of installed capacities, their total energy generation as well as the behaviour of storage units. We further study the degree of decarbonization, and at last the levelized cost of electricity (LCOE).

\subsubsection{Energy Carriers}

\begin{figure}[b]
    \centering
    \includegraphics[width=.79\textwidth]{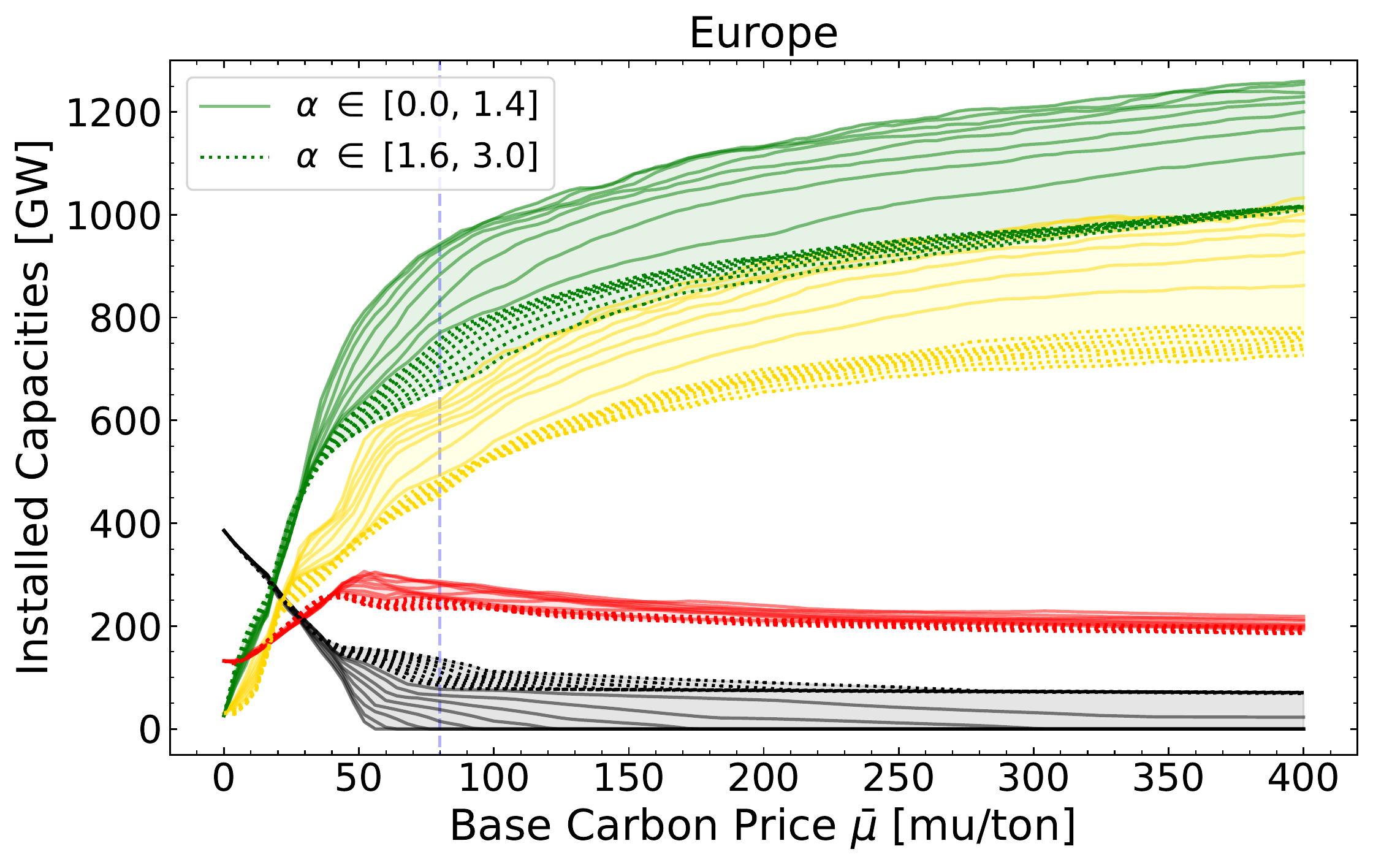}
    \caption{Optimized capacities of the different energy carrier types. The data are split into two sets of curves, where the solid lines indicate $\alpha \in [0.0, 1.4]$, and the dotted lines $\alpha \in [1.6, 3.0]$ with steps of $\Delta \alpha = 0.2$ respectively. Colors are defined as follows: green: wind, yellow: solar PV, black: coal, red: gas. A rising base carbon price disadvantages conventional energy sources. Carbon leakage happens for $\alpha \geq 1.6$.}
    \label{fig:capacities}
\end{figure}

\begin{figure}[b]
    \centering
    \includegraphics[width=.79\textwidth]{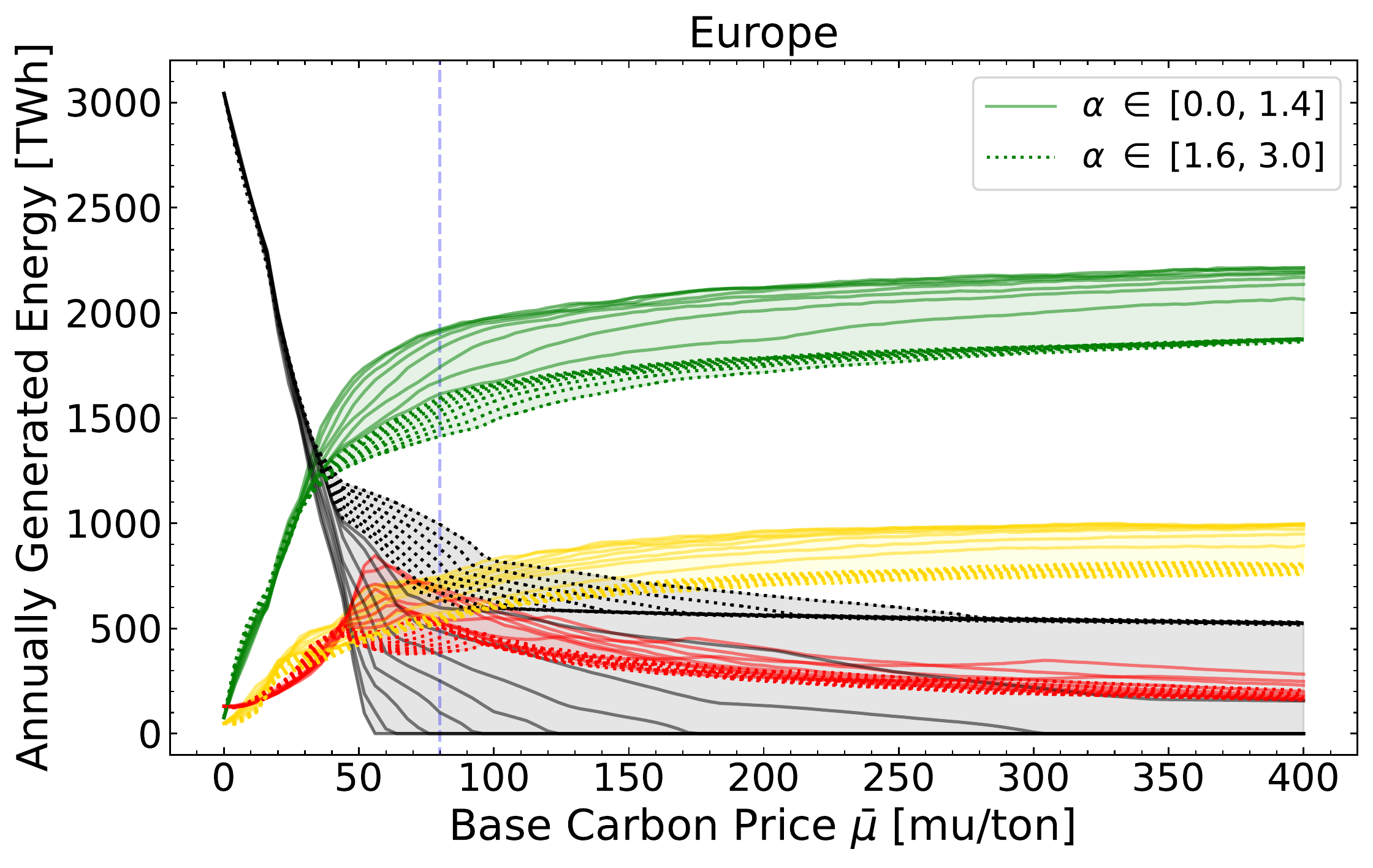}
    \caption{Energy generation from the different energy carrier types integrated over one year. The data are split into two sets of curves, where the solid lines indicate $\alpha \in [0.0, 1.4]$ and the dotted lines $\alpha \in [1.6, 3.0]$ with steps of $\Delta \alpha = 0.2$ respectively. Colors are defined as follows: green: wind, yellow: solar PV, black: coal, red: gas. The generated energy correlates with the installed capacities carrier-wise. For base carbon prices $\bar{\mu} < 50$ mu energy generation from coal is dominant. Prices above this threshold either leads to an equilibrium state ($\alpha \geq 1.6$) or an extinction ($\alpha < 1.6$) of coal power.}
    \label{fig:generation}
\end{figure}

Fig. \ref{fig:capacities} shows the optimized capacities for the different energy carrier types (green: wind, yellow: solar PV, black: coal, red: gas) as a function of the base carbon price $\bar{\mu}$ and the distribution parameter $\alpha$. Note that for an increasing base carbon price all conventional energy carriers become less competitive by definition. The carbon leakage spotted in Fig. \ref{fig:regional_analysis} does occur if the distribution parameter exceeds $\alpha \geq 1.6$. At this point the sets of curves in Fig. \ref{fig:capacities} - \ref{fig:decarbonization} split up.\\
\tab In the following we describe the main observations: First, as the base carbon price increases beyond $\bar{\mu} > 50$ mu, the renewability of the system rises as indicated by the total amount of the deployed wind (green) and solar PV (yellow) capacities. This is true for all possible choices of the distribution parameter, especially for small values $\alpha < 1.6$. Here, the effective carbon price in each region is high enough to enforce and further highly renewable system configurations. Conventional carrier types such as coal (black) and gas (red) are reduced to minimal values.\\
\tab Second, for values of $\alpha \geq 1.6$ and high base carbon prices $\bar{\mu} > 300$ mu, the system runs into equilibrium states. Here the distribution parameter does no longer strongly affect the total amount of a deployed carrier type. This is true for wind and coal capacities but less strictly so for solar PV and gas, which show a rather volatile behaviour. For small distribution parameters $\alpha < 1.6$ however, the conventional carriers do still run into a stationary equilibrium, while the renewables behave more dynamically and reveal bigger spreads of their final state.\\
\tab Third, small distribution parameters $\alpha < 1.6$ and very high base carbon prices of $\bar{\mu} > 300$ mu let coal loose its competitiveness. Hence, coal capacities do completely vanish, while gas capacities converge into a specific backup equilibrium. The opposite situation results if the effective carbon price vanishes completely. This happens in regions with rather low GDPs and for high price spreads, i.e. for large distribution parameters $\alpha \geq 1.6$. Then coal remains a competitive source of energy. Gas, in turn, serves as backup again and remains important even at high base carbon prices and very pronounced price spreads.\\
\tab Fig. \ref{fig:generation} visualises the actual amount of energy generated from all deployed carrier capacities during the time span of one year. Qualitatively, the generated energy per carrier correlates with their installed capacities respectively. It is remarkable that for base carbon prices $\bar{\mu} < 50$ mu the energy generation from coal is dominant. When the base carbon price rises and conventional sources become less competitive, coal power either runs into an equilibrium for distribution parameters $\alpha \geq 1.6$ or completely vanishes for $\alpha < 1.6$.

\subsubsection{Storage Units}

\begin{figure}[b]
    \centering
    \includegraphics[width=.79\textwidth]{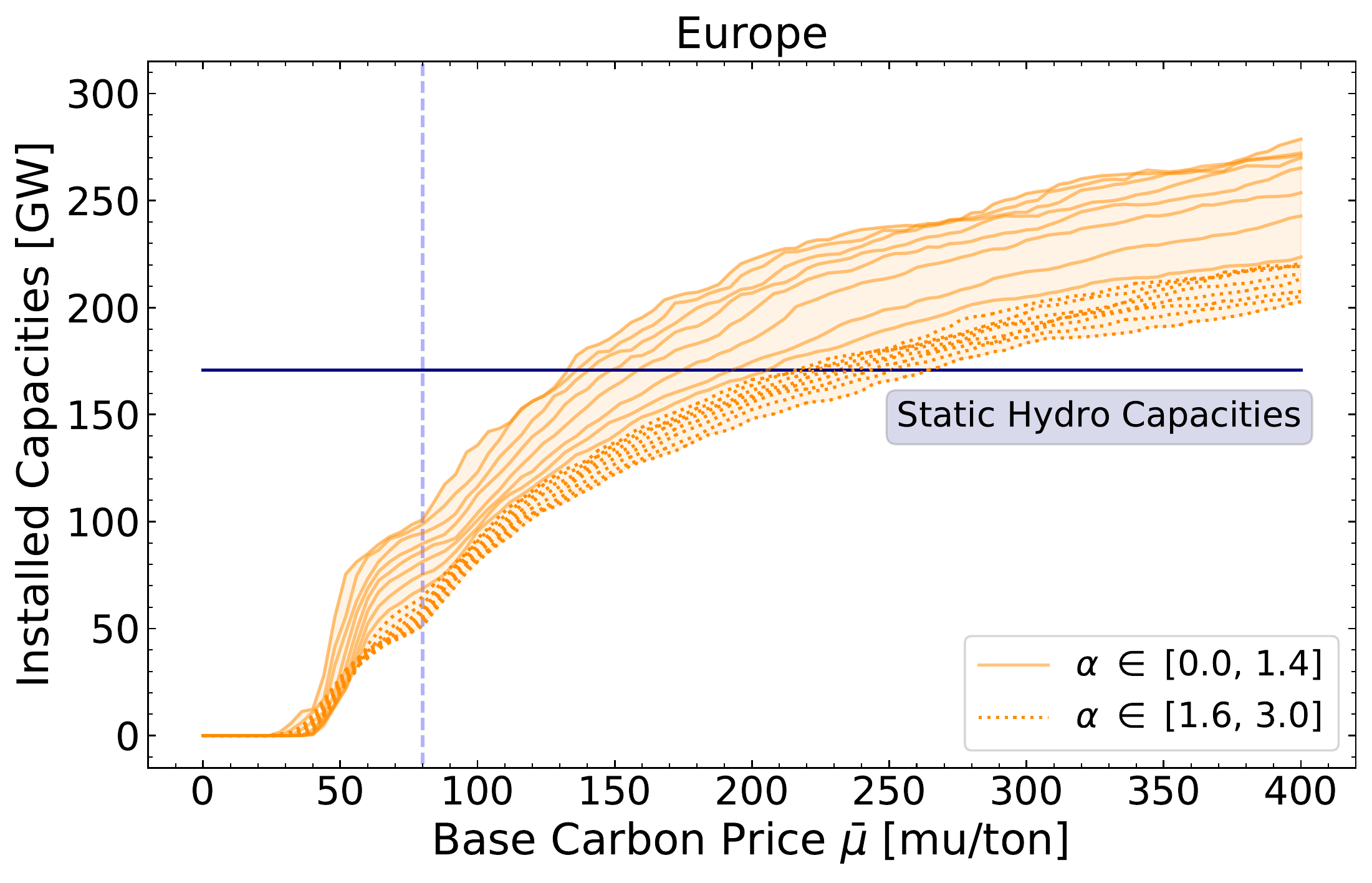}
    \caption{Optimized capacities for hydro (blue, constant) and batteries (orange). The data are split into two sets of curves, where the solid lines indicate $\alpha \in [0.0, 1.4]$ and the dotted lines $\alpha \in [1.6, 3.0]$ with steps of $\Delta \alpha = 0.2$ respectively. Battery installments grow proportionally to the renewability and their total amount is lowered by an increasing distribution parameter.}
    \label{fig:storage_capacities}
\end{figure}

\begin{figure}[b]
    \centering
    \includegraphics[width=.79\textwidth]{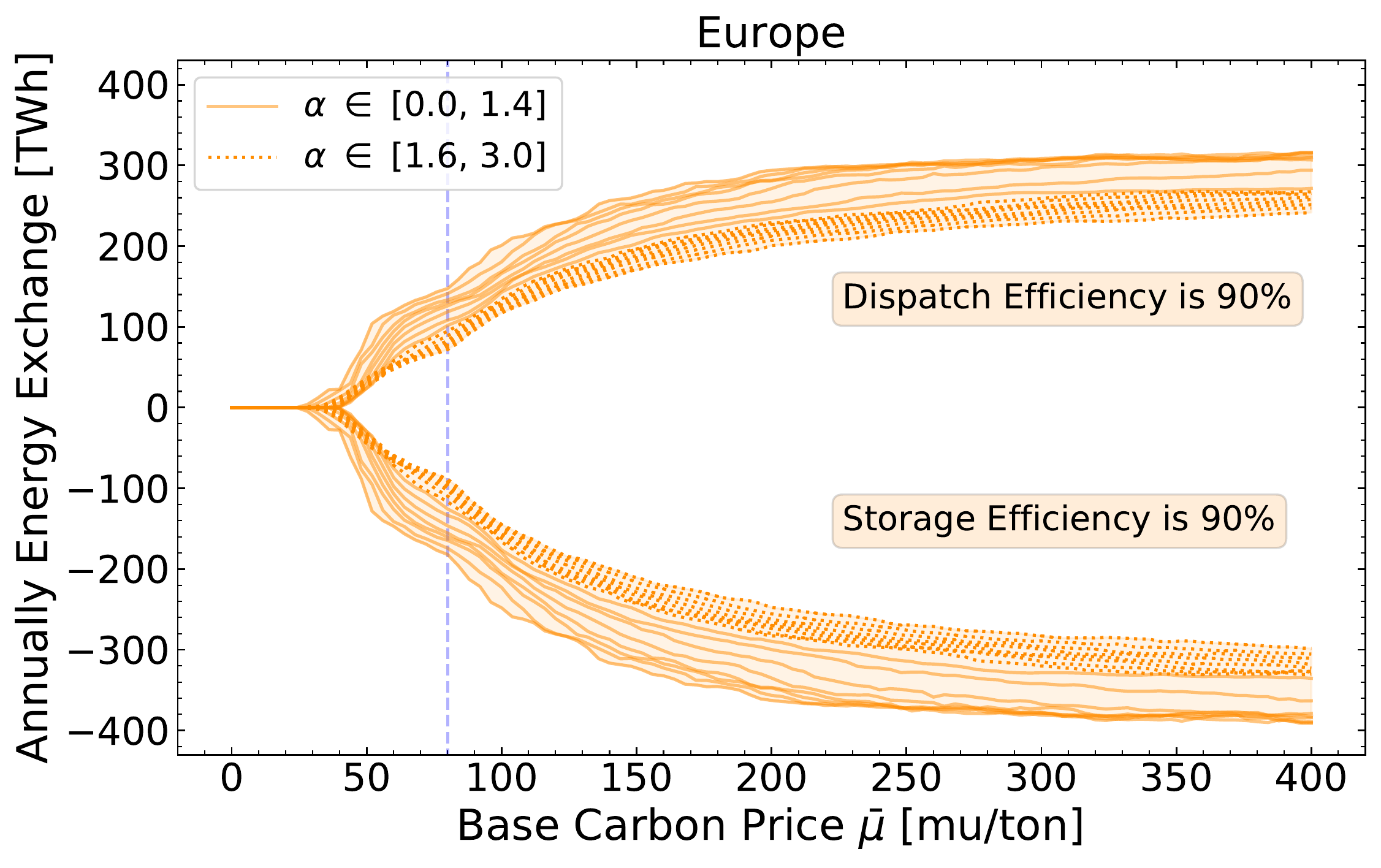}
    \caption{Storage exchange from batteries (orange) integrated over one year. The data are split into two sets of curves, where the solid lines indicate $\alpha \in [0.0, 1.4]$ and the dotted lines $\alpha \in [1.6, 3.0]$ with steps of $\Delta \alpha = 0.2$ respectively. From a base carbon price $\bar{\mu} > 50$ mu storage activity rises. An increasing distribution parameter dampens this process as the need for storage decreases together with the share of renewables in the system.}
    \label{fig:storage_exchange}
\end{figure}

Storage units play a crucial role in decarbonizing a power system. As shown in Fig. \ref{fig:storage_capacities} batteries (orange) are built up for base carbon prices $\bar{\mu} > 50$ mu. In such case the system is running into an overall renewable state as shown in Fig. \ref{fig:capacities} and \ref{fig:generation}, which is the case for any choice of the distribution parameter. Renewable power systems in turn are highly dependent on storage capabilities, since weather-driven fluctuations affect the generation from renewables and must be balanced out. Hence battery capacities grow together with the total amount of renewable capacities. However, the total amount of battery storage is limited by the distribution parameter, where a large $\alpha$ lowers the total share of renewables in the system, increases the share of conventional sources, and thus reduces the general needs for storage.\\
\tab Fig. \ref{fig:storage_exchange} shows the total energy exchange of the installed batteries, i.e. their charge and discharge behaviour integrated over one year. For base carbon prices $\bar{\mu} > 50$ mu storage activity rises as the deployed battery capacities do. However, the larger the distribution parameter, the more dampened this process, as the need for storage decreases together with the share of renewables within the system.\\
\tab Note that the electricity generation from hydro sources, despite being a fundamental part of the European power system, turns out to be negligibly small as compared to the exchange from and to batteries. For visualisation purposes it was excluded from Fig. \ref{fig:storage_exchange}.

\subsubsection{Decarbonization}

\begin{figure}[b]
    \centering
    \includegraphics[width=.79\textwidth]{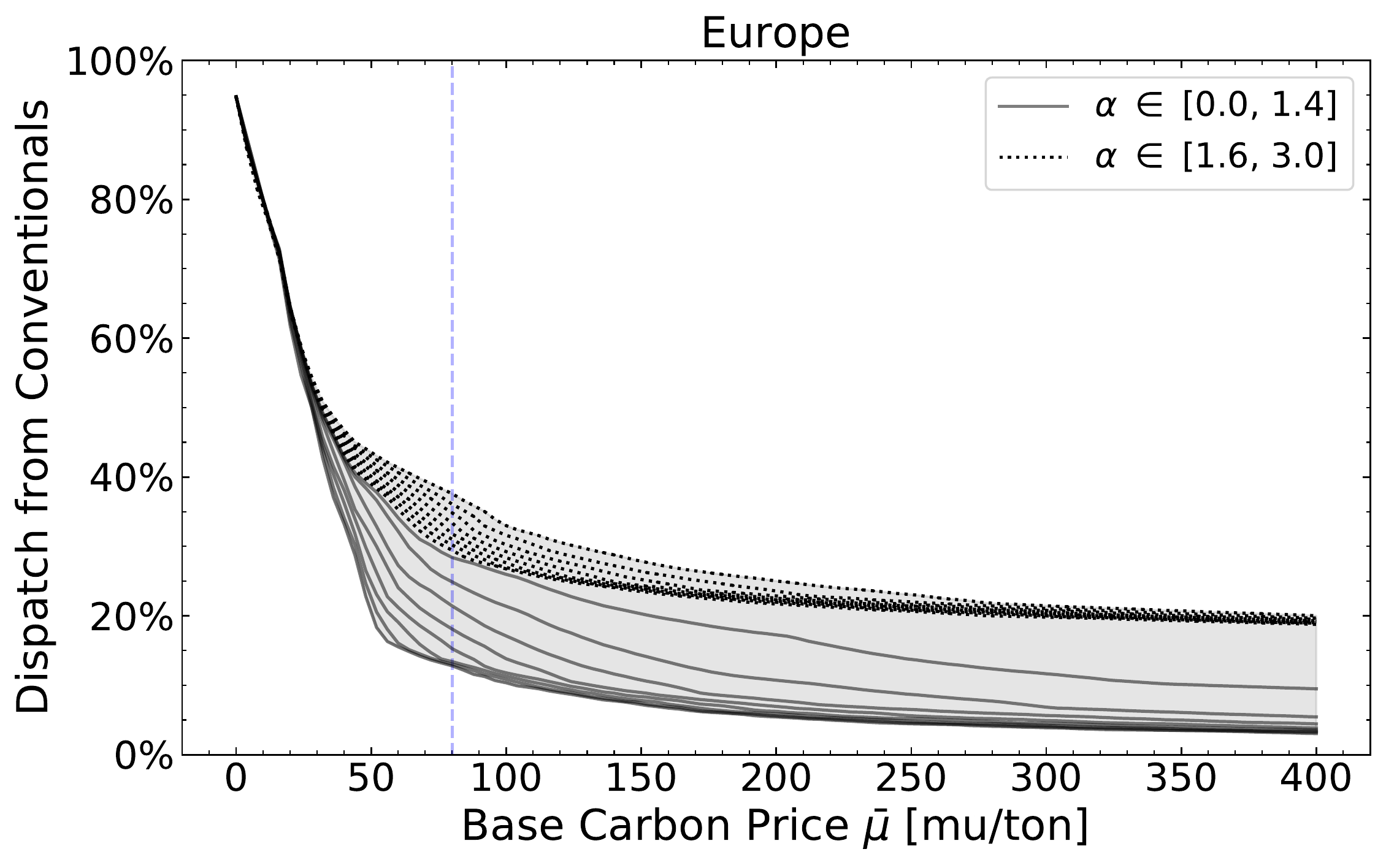}
    \caption{Total dispatch from conventional energy sources, i.e. coal and gas, in fractions of the overall generated energy. The data are split into two sets of curves, where the solid lines indicate $\alpha \in [0.0, 1.4]$ and the dotted lines $\alpha \in [1.6, 3.0]$ with steps of $\Delta \alpha = 0.2$ respectively. The solid lines indicate a systemic decarbonization process for base carbon prices $\bar{\mu} \geq 50$ mu. However, as indicated by the dotted lines, all $\alpha \geq 1.6$ run into a carbonized equilibrium state, where around 30\% of all generated energy is delivered by conventional sources.}
    \label{fig:decarbonization}
\end{figure}

Fig. \ref{fig:decarbonization} shows the total dispatch from conventional energy sources - both coal and gas - in fractions of the total electricity load as it is covered by the different carrier types and storage units. Carbon leakage happens for $\alpha \geq 1.6$, where the set of curves splits up into two subsets. For distribution parameters $\alpha < 1.6$ and base carbon prices $\bar{\mu} > 50$ mu, the solid lines indicate a systemic decarbonization process. For $\alpha \geq 1.6$, the system runs into a carbonized equilibrium state, as indicated by the dotted lines. Here around 30\% of all generated energy is dispatched by conventional energy carriers, even though very high base carbon prices $\bar{\mu} > 300$ mu are applied.\\
\tab Note that a base carbon price of $\bar{\mu} > 50$ mu already leads to a recognisable reduction of conventional generation, which at least amounts to 50\% decarbonization for $\alpha \geq 1.6$, and reaches out to nearly 90\% for $\alpha < 1.6$.

\subsubsection{Total System Cost}

\begin{figure}[b]
    \centering
    \includegraphics[width=.79\textwidth]{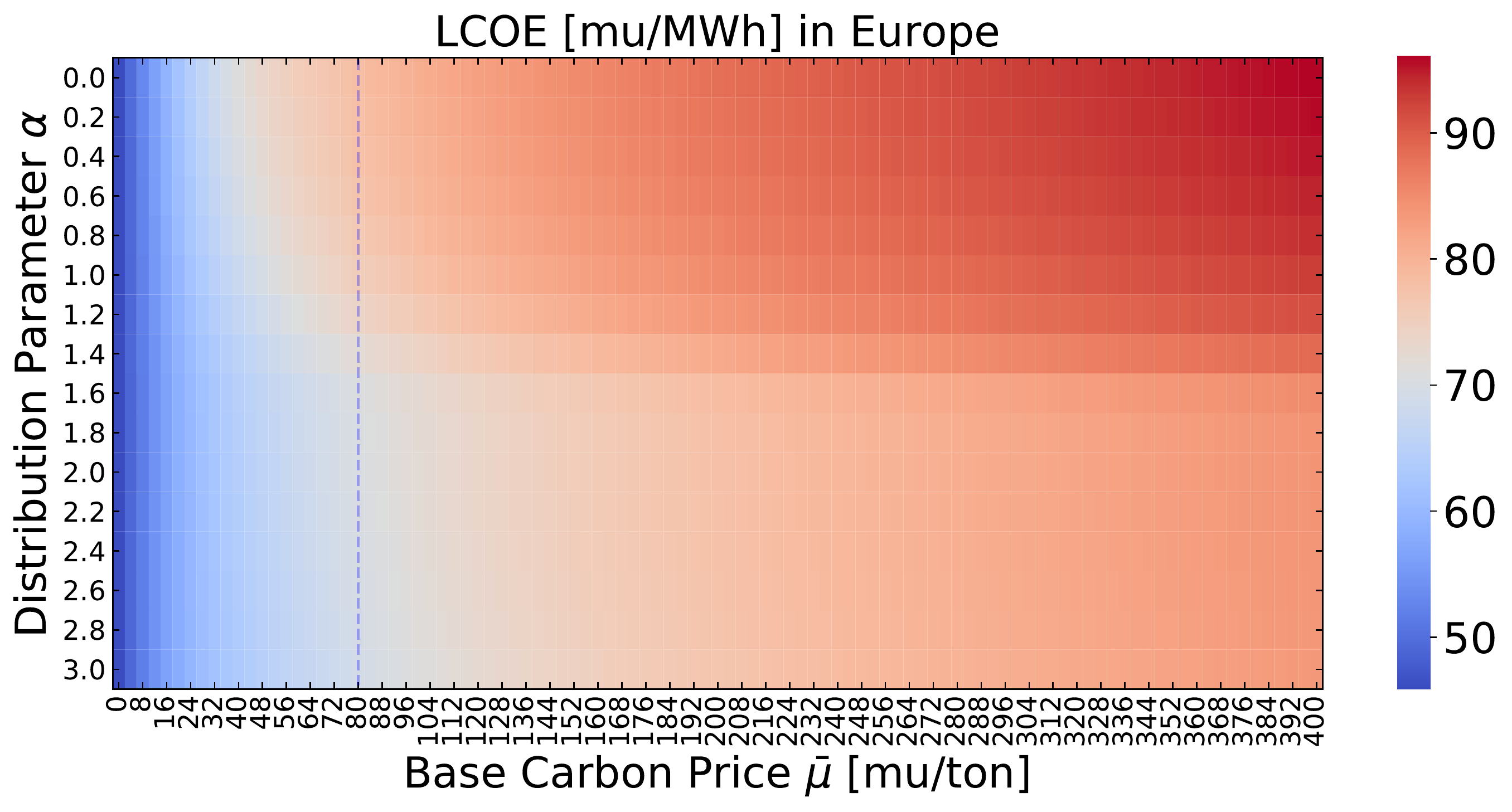}
    \caption{Levelized cost of electricity (LCOE). Note that effective carbon prices are part of the regional marginal cost for energy generation from conventional sources, i.e. coal and gas. Dark blue represents low base carbon price and low total system cost of about 50 mu/MWh. Light red visualises partially renewable states including carbon leakage in regions with low GDPs. Dark red respresents highly renewable states and total system cost up to 90 mu/MWh.}
    \label{fig:lcoe}
\end{figure}

As described in Eq. \ref{eq:minimization}, the total system cost are the main objective of the optimization problem. Fig. \ref{fig:lcoe} shows the levelized cost of electricity (LCOE) for all possible $\bar{\mu}$-$\alpha$-combinations. This LCOE color map can be categorised into three distinct fields, represented by three areas and their respective colour scale:\\
\tab First, a low base carbon price leads to low total system cost and an LCOE of about 50 mu/MWh. Then the system remains in a configuration dominated by cheap conventional energy sources. This area is colored in dark blue.\\
\tab Second, for distribution parameters $\alpha < 1.6$, the system is shifted to a highly renewable state as the base carbon price increases. This raises the LCOE up to 90 mu/MWh and more; the corresponding area is represented by dark red.\\
\tab Third, for distribution parameters $\alpha \geq 1.6$ the system is only shifted to a partially renewable state as the base carbon price increases. Then the total system cost includes carbon leakage effects in regions with rather low GDPs and the LCOE gradually increases to about 80 mu/MWh. This area is colored in light red.\\
\tab Note that the carbon leakage effect, which becomes stronger as the distribution parameter becomes larger, pushes the LCOE to lower values. Besides the price jump at $\alpha = 1.6$, where the system shifts from a highly renewable state to a more conventional one, two additional effects take place: First, LCOE-levels are shifted to the right on the abscissa, i.e. to higher base carbon prices. Second, these shifted LCOE-levels are spread within larger intervals regarding the base carbon price. Both, the shift and the spread effect are dominant for $\alpha \geq 1.6$ and base carbon prices inbetween the interval $\bar{\mu} \in [40, 120]$ mu, when carbon leakage affects the total system cost the strongest.

\section{Summary} \label{sec:sec4}

In this work, the impact of inhomogeneous carbon prices on carbon leakage was investigated.\\
\tab It was shown that unevenly distributed carbon prices within the European power system yield in high shares of conventional energy generation from coal and gas, in particular in the Baltic, Eastern and Balkans regions, where the amount of leakage is directly related to the degree of carbon price inhomogeneities.\\
\tab The quantitative dependency of carbon leakage on the distribution parameter $\alpha$ was drawn: Carbon leakage occurs for $\alpha \geq 1.6$ in the stated regions. For larger $\alpha$ also Southern Europe, here represented by the Iberian peninsula and Italy, is eventually dominated by conventional energy sources, leaving the ample potential of energy from solar PV widely unused.\\
\tab Finally, a deeper analysis of power flows within the European transmission grid revealed that a substantial fraction of energy from conventional carriers, especially coal, leaks from Eastern, Southeastern and Southern Europe to its West and North.

\subsection{Conclusions}

Regionally diversified carbon prices across Europe, as described by the scenarios developed in this paper, resemble the present reality: Carbon prices were established in more than 15 European countries from the the 1990s onwards, varying in level and coverage \cite{metcalf2020macroeconomic, carbontaxes2020}. Countries such as Sweden impose rates of around 110 EUR per ton of carbon dioxide, while in other countries such as France carbon taxes are less expensive with rates of 45 EUR per ton or lower - or are completely absent as in most countries of Eastern and Southeastern Europe.\\
\tab Considering the actual price layout, an inhomogeneous carbon price seems a plausible scenario for the future, while the results of this paper show the disastrous consequences: Carbon Leakage within the European power system and shifting of coal power plants to regions with low or vanishing carbon prices in climatically relevant orders of magnitude, where the bigger the price spread among regions, the greater the potential for this leakage.\\
\tab This has to be taken as a serious threat to the actual European emission reduction targets, underlined by the fact that the existing coal phase-out plans within the European Union - for an overview see \cite{coalexittracker} - show similar spatial patterns as presented above: While countries in Western Europe pursue ambitious coal phase-out plans, countries in Eastern Europe deny such phase-out discussions and have made clear that they will rely on coal as a substantial part of their power infrastructure.\\
\tab To curb this potential risk a joint strategy should be considered. e.g. by determining and analysing the regional cost aspects as well as by tracking and pricing carbon charged cross border flows as presented in this paper.\\
\tab We firmly emphasize: The issue of carbon leakage must be discussed early on.

\section*{Acknowledgments}
This work is financially supported by the "CoNDyNet II" project (BMBF, Fkz. 03EK3055C) and the "Net-Allok" project (BMWi, Fkz. 03ET4046A).\\
\tab We appreciate the additional support from the "EnergieSysAI" project (BMBF), the "Avicenna-Studienwerk", and the "efl - the Data Science Institute" and acknowledge the Judah M. Eisenberg - Professor Laureatus of Theoretical Physics - chair at the Fachbereich Physik at Goethe Universität, promoted by the Walter Greiner Gesellschaft zur Förderung der physikalischen Grundlagenforschung.\\
\tab We further thank Martin Greiner (Aarhus) for helpful suggestions and comments to improve this paper.

\bibliographystyle{IEEEtran}
\bibliography{references}

\end{document}